\documentclass[12pt]{article}
\usepackage[OT1]{fontenc}
\usepackage[utf8]{inputenc}
\usepackage{tocloft}
\usepackage{amsfonts}
\usepackage{amsmath}
\usepackage{amssymb}
\usepackage{array}
\usepackage{slashed}
\usepackage{bigints}
\usepackage{pifont}
\usepackage{bm}
\usepackage{bbm}
\usepackage{upgreek}
\usepackage{booktabs}
\usepackage[sort]{cite}
\usepackage{xcolor}
\usepackage{dsfont}
\usepackage{float}
\usepackage{framed}
\usepackage{graphicx}
\usepackage{indentfirst}
\usepackage{mathrsfs}
\usepackage{multirow}
\usepackage{pdflscape}
\usepackage{setspace}
\usepackage{subdepth}
\usepackage{subfig}
\usepackage{titlesec}
\usepackage{wrapfig}
\usepackage[all]{xy}
\usepackage{young}
\usepackage[vcentermath]{youngtab}
\usepackage{relsize}
\usepackage{stackengine}
\usepackage{verbatim}
\usepackage{slashed}

\newcommand{\rd}{{\rm d}}

\usepackage{hyperref}
\hypersetup{colorlinks=true}
\hypersetup{linkcolor=black}
\hypersetup{citecolor=black}
\hypersetup{urlcolor=black}

\numberwithin{equation}{section}

\usepackage[left=2.5cm,right=2.5cm,top=2.5cm,bottom=3cm]{geometry}
\fontdimen2\font=1.2\fontdimen2{\jot}{5pt}

\newlength{\bibitemsep}
\newlength{\bibparskip}
\let\oldthebibliography\thebibliography
\renewcommand\thebibliography[1]{%
  \oldthebibliography{#1}%
  \setlength{\parskip}{\bibitemsep}%
  \setlength{\itemsep}{\bibparskip}%
}

\titleformat{\section}{\bfseries}{\thesection.}{4pt}{}
\titlespacing{\section}{0pt}{20pt}{6pt}

\titleformat{\subsection}{\normalfont\itshape}{\thesubsection.}{4pt}{}
\titlespacing{\subsection}{0pt}{15pt}{6pt}

\titleformat{\subsubsection}{\normalfont\itshape}{\thesubsubsection.}{4pt}{}
\titlespacing{\subsubsection}{0pt}{15pt}{6pt}

\titleformat{\paragraph}{\normalfont\itshape}{\theparagraph.}{4pt}{}
\titlespacing{\paragraph}{0pt}{15pt}{6pt}

\renewcommand{\tilde}{\widetilde}

\newcommand{\half}{\frac{1}{2}}

\DeclareMathOperator{\Tr}{Tr}
\DeclareMathOperator{\lcm}{lcm}
\DeclareMathAlphabet{\mathbfsf}{OT1}{cmss}{bx}{n}

\newcommand{\Z}{{\mathbb Z}}
\newcommand{\C}{{\mathbb C}}
\newcommand{\R}{{\mathbb R}}

\newcommand{\mc}[1]{\mathcal{#1}}

\newcommand{\mf}[1]{\mathfrak{#1}}

\newcommand{\SL}{\mathscr{L}}

\newcommand{\bZ}{\mathbb{Z}}
\newcommand{\bC}{\mathbb{C}}

\newcommand{\cA}{\mathcal{A}}
\newcommand{\cB}{\mathcal{B}}
\newcommand{\cC}{\mathcal{C}}
\newcommand{\cD}{\mathcal{D}}
\newcommand{\cE}{\mathcal{E}}

\newcommand{\cG}{\mathcal{G}}

\newcommand{\cM}{\mathcal M}
\newcommand{\cN}{\mathcal{N}}
\newcommand{\cO}{\mathcal{O}}

\newcommand{\cP}{\mathcal{P}}

\newcommand{\cY}{\mathcal{Y}}

\newcommand{\U}{\mathrm{U}}

\newcommand{\ed}{\,.}
\newcommand{\ec}{\,,}

\newcommand{\be}{\begin{equation}}
\newcommand{\ee}{\end{equation}}
\newcommand{\beq}{\begin{equation}}
\newcommand{\eeq}{\end{equation}}

\newcommand{\zero}{^{(0)}}
\newcommand{\meno}{^{(-1)}}
\newcommand{\one}{^{(1)}}

\newcommand{\YM}{\textrm{YM}}

\newcommand{\ii}{{\rm i}}
\newcommand{\e}{{\rm e}}
\newcommand{\eff}{\textrm{eff}}

\allowdisplaybreaks 

\DeclareFontShape{OT1}{cmr}{mx}{n}%
{<->cmr10}{}
\newcommand{\mytitlefont}{\fontseries{mx}\selectfont}
\DeclareMathAlphabet{\titlemath}{OT1}{cmr}{mx}{n}

\begin{document}

%
\begin{titlepage}
\begin{center}
~\\[2cm]
{\fontsize{29pt}{0pt} \mytitlefont Instantons, Symmetries and\\[0.1cm] Anomalies in Five Dimensions}
~\\[1.25cm]
Pietro Benetti Genolini\,$^1$ and Luigi Tizzano\,$^2$\hskip1pt
~\\[0.5cm]
{$^1$~{\it Department of Applied Mathematics and Theoretical Physics,\\
University of Cambridge, Wilberforce Road, Cambridge, CB3 OWA, UK}}
~\\[0.15cm]
{$^2$~{\it Simons Center for Geometry and Physics, SUNY, Stony Brook, NY 11794, USA}}
~\\[1.25cm]
			
\end{center}
\noindent 
All five-dimensional non-abelian gauge theories have a $U(1)_I$ global symmetry associated with instantonic particles.  We describe an obstruction to coupling $U(1)_I$ to a classical background gauge field that occurs whenever the theory has a one-form center symmetry. This is a finite-order mixed 't Hooft anomaly between the two symmetries. We also show that a similar obstruction takes place in gauge theories with fundamental matter by studying twisted bundles for the ordinary flavor symmetry. We explore some general dynamical properties of the candidate phases implied by the anomaly. Finally, we apply our results to supersymmetric gauge theories in five dimensions and analyze the symmetry enhancement patterns occurring at their conjectured RG fixed points.

\vfill 
\begin{flushleft}
September 2020
\end{flushleft}
\end{titlepage}
%
		
	
\setcounter{tocdepth}{3}
\renewcommand{\cfttoctitlefont}{\large\bfseries}
\renewcommand{\cftsecaftersnum}{.}
\renewcommand{\cftsubsecaftersnum}{.}
\renewcommand{\cftsubsubsecaftersnum}{.}
\renewcommand{\cftdotsep}{6}
\renewcommand\contentsname{\centerline{Contents}}
	
\tableofcontents


\section{Introduction and Summary}

Gauge theories in $d>4$ are infrared free and have a Landau pole singularity in the ultraviolet requiring a cutoff regulator. They should be regarded only as IR effective theories. Nevertheless, it is interesting to explore what kind of dynamical phenomena could arise in these models. 

Here we will focus solely on the properties of gauge theories in five dimensions. A unique feature of these gauge theories is that they carry a $U(1)\zero_I$ conserved current associated to Yang--Mills instanton configurations, which in five-dimensional spacetime appear as particles.\footnote{Throughout, we use a superscript $(q)$ in parentheses to indicate a $q$-form symmetry group.} These particles carry a charge $Q_I$ (the instanton number) and are created by acting with \emph{instanton operators} on the vacuum.  More abstractly, instanton operators can be thought of as disorder operators (similarly to 't Hooft lines in 4$d$ gauge theories) obtained by imposing boundary conditions on the gauge fields at the insertion point \cite{Lambert:2014jna, Tachikawa:2015mha}. These are completely analogous to the more familiar example of monopole operators in three-dimensional gauge theories \cite{Borokhov:2002ib, Borokhov:2002cg}.

Five-dimensional Yang--Mills theory with $G= SU(N)$ also possesses a generalized one-form global symmetry $\bZ\one_N$, associated to the center of the gauge group, whose effects can be studied in detail by activating a corresponding nontrivial background two-form $\bZ_N$ gauge field $\cB$ \cite{Kapustin:2014gua, Gaiotto:2014kfa}. The effect of $\cB$ is to implement the 't Hooft twisted boundary conditions \cite{tHooft:1979rtg}, showing that we are considering a bundle whose structure group is not simply connected. In this case, the instanton number $Q_I$ can be fractional and the $U(1)\zero_I$ symmetry is destroyed.

In this note, we describe how the $U(1)\zero_I$ global symmetry and the $\bZ\one_N$ generalized one-form global symmetry participate in a mixed 't Hooft anomaly. In fact, under a background large gauge transformation for the $U(1)\zero_I$ symmetry, the partition function picks up an additional nontrivial phase factor depending on $\cB$. More precisely, the fractional part of the instanton number is measured by a cohomological class constructed from the $\bZ_N$ background field. Such transformation law means that the partition function of the theory is not a complex number but rather a section of a line bundle, which is the signal of an anomaly. Our analysis will be close in spirit to the work \cite{Komargodski:2017dmc} which has studied the same anomaly in the three-dimensional Abelian Higgs model.\footnote{See also \cite{Bergman:2020ifi} for a recent application in ABJM-type theories. A recent paper \cite{Apruzzi:2020zot} has focused on the interplay between $U(1)_I\one$ and $\Z_N\one$ for six-dimensional gauge theories, with an interest towards $(1,0)$ SCFTs. The $U(1)_I\one$ symmetry is always gauged in this case, leading to further constraints on the analysis of 't Hooft anomaly matching in $6d$ SCFTs \cite{Cordova:2016emh, Cordova:2020tij}.}

Anomalies in $d$ dimensions are often described in terms of an inflow mechanism from $(d+1)$-dimensional classical local functionals, the so-called anomaly theories, of background fields. Anomaly theories have a gauge invariant action on closed manifolds, whereas on a space with boundary their variation under a background gauge transformation cancels the anomaly of the original theory.\footnote{They are $(d+1)$-dimensional invertible field theories, as defined in \cite{Freed:2004yc, Freed:2012bs}.}

Since 't Hooft anomalies must match along any RG flow \cite{tHooft:1979rat}, their presence is very useful to put constraints on the phases of the theory. Whilst this has been used extensively in four and lower dimensions, the same argument also applies to field theories in higher dimensions (see for instance an early application in six dimensions in \cite{Intriligator:2000eq}). 

Let us now briefly summarize our findings in the case of $5d$ $SU(N)$ Yang--Mills theory which, in this context, we always view as an IR effective theory regulated by a UV cutoff $\Lambda_5$. We first analyze the behavior of the theory upon reducing on a circle --- or equivalently at finite temperature. An important phenomenon that takes place here is that the mixed anomaly between $U(1)\zero_I$ and $\bZ\one_N$ persists at finite temperature indicating that the theory should be in an ordered phase for any value of the temperature. This special behavior is due to the presence of higher-form symmetries, as first emphasized in \cite{Gaiotto:2017yup}. We consider some qualitative features of the small and large temperature limit and find that a consistent candidate phase at small temperature implies that the higher form $\bZ\one_N$ should be spontaneously broken. However, without additional knowledge of the UV behavior and symmetries realized by the dynamics of the theory it is difficult to exclude other candidate phases involving spontaneous breaking of $U(1)\zero_I$ or more exotic possibilities such as symmetry-preserving non-trivial RG fixed points. 

Supersymmetry gives us further insight about the symmetries that should be realized at high energies. Indeed, it was suggested in  \cite{Seiberg:1996bd, Douglas:1996xp, Morrison:1996xf} that certain $5d$ SUSY gauge theories have a non-trivial supersymmetric RG fixed point. If we flow out of the RG fixed point by activating a supersymmetric mass deformation, the low-energy effective theory is a weakly-coupled $5d$ gauge theory to which our analysis can be applied. For this reason, we extend our results to discuss candidate supersymmetric phases saturating the mixed anomaly between $U(1)\zero_I$ and $\bZ\one_N$. For gauge theories with fundamental matter and global flavor group $G\zero_F$, which do not have a one-form symmetry $\bZ\one_N$, we show that there still is a finite-order mixed anomaly (i.e. its coefficient only takes a finite number of values) involving the instanton symmetry $U(1)\zero_I$ and $G\zero_F$.

In the case of pure $SU(2)$ SYM theory, we find that the simplest possibility among the candidate supersymmetric phases is that the UV the theory is in a gapless deconfined phase which preserves the $U(1)\zero_I$ symmetry and spontaneously breaks the $\bZ\one_N$ symmetry. Such phase is perfectly consistent with the proposed UV fixed point for this theory, known as $E_1$, characterized by global symmetry enhancement to $SU(2)\zero_I$. 

The mixed anomaly we describe is also compatible with a gapless deconfined phase which preserves both $U(1)\zero_I$ and $\bZ\one_N$. If such phase is realized dynamically, we can further ask if it is possible to express the anomaly in terms of UV symmetries. Interestingly, it turns out that this last requirement is only mathematically compatible with $SO(3)\zero_I$ UV symmetry enhancement of the low-energy $U(1)\zero_I$ instanton symmetry. It would be interesting to further analyze this phase by computing protected observables that can probe the topology of the enhanced global symmetry group.

For theories with matter in the fundamental representation a richer picture is available and we also discuss a candidate phase characterized by spontaneous breaking of the $U(1)\zero_I$ symmetry. 

The mixed anomaly we present here should also be useful for the study of dynamical interfaces, which we leave for future work. Namely, we can vary the profile of a background field for the $U(1)\zero_I$ symmetry along one of the spacetime dimensions and use the anomaly to put constraints on the resulting worldvolume dynamics. For supersymmetric theories, it would be nice to make contact with the analysis of \cite{Gaiotto:2015una}.

\medskip

The paper is organized as follows. In Section \ref{sec:pureYM}, we focus on $SU(N)$ Yang--Mills theory, showing that there is a mixed 't Hooft anomaly between the $U(1)\zero_I$ symmetry and $\Z\one_N$. We also show that $\Z\one_N$ can be explicitly broken to $\Z\one_{\gcd(N,k)}$ by introducing a level-$k$ $5d$ Chern--Simons term. We study the finite temperature behavior of the five-dimensional theory in some limits and present more general implications of the anomaly for its dynamics. We also describe the relation with the anomaly of four-dimensional Yang--Mills at $\theta=\pi$ found in \cite{Gaiotto:2017yup}. In Section \ref{sec:SUSYtheories}, we comment on how to extend the mixed anomaly to supersymmetric theories. In pure SYM, we still have the same mixed anomaly between $U(1)\zero_I$ and the center symmetry, since the fields that have to be added to pure Yang--Mills transform in the adjoint representation of the gauge group and thus do not break the center symmetry. This is not the case if we add hypermultiplets transforming in arbitrary representations of the gauge group. However, we argue that it is still possible to have a mixed 't Hooft anomaly between $U(1)\zero_I$ and the flavor symmetry, and we show in detail its presence in the case of $SU(2)$ gauge theory with $N_f$ hypermultiplets in the fundamental, with flavor symmetry $Spin(2N_f)$. We conclude by presenting some remarks on the implications of the anomaly for the UV fixed point. The results presented in Appendix \ref{app:5dQCD} do not rely on supersymmetry and give a general description of how to compute the mixed anomaly in presence of fundamental matter.

\section{Yang--Mills Theory in Five Dimensions}
\label{sec:pureYM}

\subsection{Global Symmetries and Background Fields}
\label{subsec:CenterSymm}

Consider the Lagrangian of five-dimensional pure Yang--Mills theory with gauge group $SU(N)$ on a Euclidean spacetime $\cM_5$
\be\label{eq:YM}
\SL_{\YM} = \frac{1}{g_5^2} \Tr F \wedge \star F \ed
\ee
The theory has an ordinary $U(1)\zero_I$ global symmetry whose conserved current is given by
\be
\label{eq:JI}
J_I = \frac{1}{8\pi^2} \star \Tr F \wedge F \ed
\ee
The corresponding charge is given by an integral
\be
\label{eq:QI}
Q_I (\Sigma_4) = \int_{\Sigma_4} \star J_I = \frac{1}{8\pi^2}\int_{\Sigma_4} \Tr F \wedge F \in \bZ \ed
\ee
 where $\Sigma_4$ is a codimension $1$ surface. Thus, the state charged under the symmetry is a solitonic configuration which is the uplift of four-dimensional Yang--Mills instantons, and which in the vacuum of the five-dimensional gauge theory appears as an instantonic particle.\footnote{By construction, we are considering $SU(N)$ gauge bundles $\cE$, so $J_I = -\star c_2(\cE)$ where $c_2$ refers to the second Chern class of $\cE$, and $Q_I$ is integer since $c_2\in H^4(BSU(N),\bZ)$ is an integral class.} The $U(1)\zero_I$ symmetry is often referred to as \textit{instanton symmetry} because of the nature of the charged operators. In addition, the theory has a one-form global symmetry $\bZ\one_N$ associated to the center of the gauge group \cite{Gaiotto:2014kfa}. The one-form symmetry acts on fundamental Wilson loops $W_F$ as: $W_F \to \e^{\frac{2\pi \ii}{N}} W_F$.

These two symmetries will play a very important role in this paper. In order to keep track of the $U(1)\zero_I$ symmetry, we introduce a background gauge field denoted by $\cA$ which couples to $J_I$ as
\be\label{eq:coupling}
\delta \SL = \ii \, \cA \wedge \star J_I \ed
\ee

Suppose that $\cM_5$ is such that we can perform a large gauge transformation for the $U(1)\zero_I$ background gauge field: $\cA \to \cA + \lambda\one$ with $\lambda\one$ being a closed but non-exact one-form. 
By construction, $\lambda\one$ has winding $2\pi \ell \in2\pi\Z$ around the non-trivial one-cycle, and there exists a non-trivial four-cycle $\Sigma_4$ by Poincaré duality (for concreteness, one could consider $\cM_5 = S^1\times \Sigma_4$). As a function of the background field $\cA$, the partition function is subject to the transformation law
\be\label{eq:largegauge}
Z[\cA] \to Z[\cA]\exp\left(2\pi \ii \ell \, Q_I(\Sigma_4) \right)\ed
\ee
Clearly, since $\ell$ and $Q_I(\Sigma_4)$ are integers, both the partition function and effective action are invariant under large gauge transformations. In the following we will restrict to $\cM_5$ being spin, as we are interested in the possibility of adding spinor fields.

We then activate a background $\bZ_N$ gauge field for the center symmetry, and denote it by $\cB \in H^2(\cM_5,\Z_N)$. One way to think about this operation is that we are including in the path integral $PSU(N)$ bundles $\cE$ that are not $SU(N)$ bundles \cite{Gukov:2013zka, Kapustin:2013uxa, Kapustin:2014gua}, i.e. there is a non-trivial Brauer class $w\in H^2(BPSU(N),\Z_N)$, and we are setting $\cB = w$.\footnote{A $PSU(N)$ bundle may have an obstruction to lifting it to a $SU(N)$ bundle. Many physics papers on discrete anomalies have been referring to $w$ as second Stiefel--Whitney class of $\cE$. However, this mathematical terminology is typically adopted only for $SO(N)$ bundles. The Stiefel--Whitney class for an $SO(3)$ bundle corresponds to the Brauer class for a $PSU(2)$ bundle.}
However, crucially, $PSU(N)$ bundles do not necessarily have integer instanton number, since $\star J_I$ is not anymore an integral class. Indeed, the integral \eqref{eq:QI} computed using the $PSU(N)$ curvature form gives $Q_I(\Sigma_4) \in \frac{1}{N}\Z$.\footnote{On a non-spin manifold $\Sigma_4$, we would have $Q_I(\Sigma_4) = k'/2N \mod 1$ (for a generic group, this would involve the dual Coxeter number) \cite{Vafa:1994tf, Witten:2000nv, Aharony:2013hda}.}

For even $N$, the fractional part of $Q_I(\Sigma_4)$ can be completely expressed in terms of the background field $\cB$
\be
\label{eq:FractionalPartSUN}
\int_{\Sigma_4}\left( \frac{1}{8\pi^2}\Tr F\wedge F  + \frac{1}{2N}\cP(\cB) \right) \ \in \ \bZ \ec
\ee
where $\cP(\cB) \in H^4(\cM_5, \Z_{2N})$ is the Pontryagin square whose reduction modulo $N$ is $\cB\cup \cB$.\footnote{For odd $N$, we write $\cP(\cB)$ meaning $\cB \cup \cB \in H^4(\cM_5, \bZ_N)$, with a common abuse of notation.} 
 
We thus deduce that when the minimal coupling \eqref{eq:coupling} is present, a large gauge transformation for the $U(1)_I\zero$ background field induces a non-trivial change of the partition function
\be
\label{eq:LargeGaugeBv2}
Z[\cA,\cB] \to Z[\cA,\cB]\exp \left( -\frac{2\pi\ii}{2N} \ell \int_{\Sigma_4}\cP(\cB) \right) \ed
\ee

The partition function, as a function of the background gauge fields, is no longer invariant under the global $U(1)\zero_I$. Such non-invariance is due to the presence of the background field $\cB$ which implements ’t Hooft twisted boundary conditions and as such modifies the instanton number quantization. Since there is no additional five-dimensional counterterm that can be added to cancel the variation \eqref{eq:LargeGaugeBv2}, this implies that five-dimensional pure Yang--Mills theory has a mixed 't Hooft anomaly between $U(1)\zero_I$ and $\bZ\one_N$.

The mixed 't Hooft anomaly can only be cancelled by a six-dimensional anomaly theory. In our case, the relevant partition function reads
\be\label{eq:6danomaly}
A_6[\cA, \cB] = \exp\left( \frac{2\pi \ii}{2N}\int_{\cY_6} \frac{\rd \cA}{2\pi}  \,  \cP(\cB) \right) \ed 
\ee
Whilst this is well-defined on a closed space, if instead $\partial \cY_6= \cM_5$, its gauge variation cancels the anomalous variation \eqref{eq:LargeGaugeBv2} via inflow, so that the coupled theory $Z[\cA,\cB]A_6[\cA,\cB]$ is anomaly-free.

In this note, we study what constraints are imposed on the dynamics of five-dimensional Yang--Mills theory as a result of the mixed 't Hooft anomaly \eqref{eq:6danomaly}.

Note that the arguments in this section can be easily extended to gauge groups that are not $SU(N)$ using the results of \cite{Witten:2000nv}. For instance, $Sp(N)$ Yang--Mills theory has a $\Z_2\one$ center symmetry for which we can turn on a background gauge field $\cB\in H^2(\cM_5, \Z_2)$ that is identified with the obstruction to lifting a $Sp(N)/\Z_2$ bundle to a $Sp(N)$ bundle.\footnote{We also recall that in five dimensions there is an additional discrete term that identifies the bundles that we sum over \cite{Douglas:1996xp}. The discrete term is associated to $\pi_4(G)$, which is trivial for any simply connected compact Lie group apart from $Sp(N)$, for which we have $\pi_4(Sp(N)) \cong \bZ_2$. Such $\bZ_2$ is not a symmetry but rather a choice of $Sp(N)$ Yang--Mills theory.\label{footnote:discrete}} As in the $SU(N)$ case, the instanton number \eqref{eq:QI} may be non-integer -- on spin manifolds, for odd $N$ -- and this in turn implies that under a large gauge transformation for $U(1)_I\zero$ the partition function picks up the factor
\be
\label{eq:SpNAnomaly}
Z[\cA,\cB] \to Z[\cA,\cB] \exp \left( \frac{2\pi \ii N}{4} \ell \int_{\Sigma_4}\mc{P}(\cB) \right) \, .
\ee
In this case, we notice that this is trivial for even $N$, since the integral of the Pontryagin square would be even on a spin manifold. Finally, it is interesting to ask what happens to the mixed anomaly once we take into account the effect of explicitly breaking the symmetry $U(1)\zero_I$. Consider summing over instanton configurations with charges multiple of $m$ and thus explicitly breaking $U(1)\zero_I$ down to $\bZ_m\zero$.  We can again activate a background gauge field $\cA^{m}$ for $\bZ_m\zero$ which is related to the $U(1)\zero_I$ background gauge field by \cite{Komargodski:2017dmc}
\be
\exp\left(2\pi \ii \frac{1}{m} \int \cA^m\right) = \exp\left(\ii \int \cA\right) \ed
\ee
When $N$ is even, the anomalous variation of the partition function is given by\footnote{It can be easily shown that similar conclusions will hold true also for odd values of $N$.}
\be\label{anomaz}
Z[\cA^m,\cB] \to Z[\cA^m,\cB]\exp \left( -\frac{2\pi\ii}{2N} \ell \int_{\Sigma_4}\cP(\cB) \right) \ec
\ee
which could be cancelled by a five-dimensional counterterm of the form
\be
2\pi \ii \frac{p}{2N} \int \cA^m \cP(B) \ed
\ee
The presence of such counterterm modifies the anomalous variation \eqref{anomaz}:
\be
Z[\cA^m,\cB] \to Z[\cA^m,\cB]\exp \left( -2\pi\ii\ell\,\frac{mp -1 }{2N}  \int_{\Sigma_4}\cP(\cB) \right)\ec
\ee
which is nontrivial if and only if $\gcd(N,m)\neq 1$. We thus conclude that the mixed anomaly \eqref{eq:LargeGaugeBv2} is robust under explicit symmetry breaking of $\U(1)\zero_I$ preserving a $\bZ_m\zero$. This phenomenon is particularly interesting if we are interested in regulating the short distance physics by placing the theory on a lattice which often induces such explicit symmetry breaking effects.

\subsection{Chern--Simons Terms}
\label{subsec:YMCS}

The Lagrangian \eqref{eq:YM} admits a level-$k$ five-dimensional Chern--Simons term given by
\be
\label{eq:cs5}
\SL_{\rm CS} = -\frac{\ii k}{24 \pi^2}\Tr\left(A \wedge F \wedge F + \frac{\ii}{2} A \wedge A \wedge A \wedge F - \frac{1}{10} A \wedge A \wedge A \wedge A \wedge A\right)\ec
\ee
where $k \in \bZ$. This is proportional to the totally symmetric tensor denoted by $d_{abc} \equiv \half \Tr t_a\{t_b, t_c\}$, with $t_a$'s being the gauge algebra generators. Since we are studying a $SU(N)$ theory, $d_{abc}$ is only non-vanishing for $N \geq 3$.

The presence of a Chern--Simons interaction modifies the one-form symmetry of the original theory. This phenomenon has been recently studied also in \cite{Morrison:2020ool, Albertini:2020mdx, Bhardwaj:2020phs}  with different techniques. Let us introduce a single adjoint scalar field to Higgs the $SU(N)$ gauge group. At low energy, the theory has a $U(1)^{N-1}$ gauge group whose fields $a^i$ are given in terms of the following decomposition 
\be
A = \sum^{N-1}_{i= 1 }a^i T_i\ec
\ee
where $T_i$'s are Cartan generators of $SU(N)$. As emphasized in \cite{Cordova:2019jnf}, adjoint Higgsing of pure Yang--Mills theory gives rise to a spontaneously broken $\big(U(1)\one\big)^{N-1} \times \big(U(1)\one\big)^{N-1}$ one-form symmetry. This symmetry is only accidental, as the UV one-form symmetry is still $\bZ\one_N$ which acts on low energy gauge fields shifting them by a flat $\Z_N$ connection $\epsilon$
\be
\label{eq:Shift}
a^j \to a^j + \frac{2\pi j}{N}\epsilon\ed
\ee

Let us now analyze how the Chern--Simons term in the UV Lagrangian \eqref{eq:YM} modifies this argument. Whenever the above transformation is applied to \eqref{eq:cs5}, there are mixed terms of the form
\be
\SL_{\rm CS} \to \SL_{\rm CS}  -2\pi\ii \sum_{q,r} \frac{k_{qr}}{24 \pi^2 N}\, \epsilon \wedge \rd a^q \wedge \rd a^r + \dots\,\,\ed
\ee
Thus, the UV  $\bZ\one_N$ symmetry cannot be fully preserved without spoiling the quantization condition of $k$ in the Chern--Simons Lagrangian. This implies that five-dimensional $SU(N)_k$ theories have a discrete one-form symmetry given by $\bZ\one_{\gcd(N,k)}$. Upon Higgsing, the gauge fields now transform as
\be
a^j \to a^j + \frac{2\pi j}{\gcd{(N,k)}}\epsilon\ec
\ee
in such a way that the low-energy Lagrangian is well-defined.

\subsection{Finite Temperature Analysis}
\label{Temperaturev2}
The mixed 't Hooft anomaly \eqref{eq:LargeGaugeBv2} has an interesting connection to four-dimensional physics. To see this, we study the five-dimensional theory on a circle of radius $\beta$ by looking at the background $\cM_5 = \cM_4 \times S^1_\beta$. This is equivalent to considering five-dimensional Yang--Mills theory at a finite temperature $T = \frac{1}{\beta}$. In this section, we will perform two different reductions of pure $SU(N)$ Yang--Mills theory. We will first perform a dimensional reduction on the circle, corresponding to the standard thermal ensemble. Then,  we will also introduce a chemical potential $\mu$ for the instanton symmetry $U(1)\zero_I$ and look at the ensemble $e^{-\beta H - \mu Q_I}$.

Consider first the effects of the dimensional reduction using the Kaluza--Klein ansatz. The effective theory on a circle has a compact adjoint scalar $\Phi$ which comes from the component of the dynamical gauge field along $S^1_\beta$. Moreover, we decompose the $U(1)\zero_I$ background gauge field $\cA$ as 
\be
\cA = \tilde{\cA}_i \,\rd x^i + \frac{a}{2\pi\beta} \, \rd\psi \ec
\ee
where $x^i$ are local coordinates on $\cM_4$, $\psi \sim \psi + 2\pi \beta$ is the coordinate along $S^1_\beta$ and $a$ is a compact scalar field on $\cM_4$. From the above, it follows that \eqref{eq:coupling} 
becomes
\beq\label{axion}
\frac{\ii}{8\pi^2}\int a \, \Tr F \wedge F + \frac{\ii \beta}{2\pi}\int \tilde{\cA} \wedge \Tr (F \wedge \cD \Phi)  \ed
\eeq
The first term is an axion coupling, and the periodicity of $a$ can be informally related to the presence of a ``$(-1)$-form symmetry'' which we denote here by $U(1)\meno_I$. Note that $U(1)\meno_I$ is not a real global symmetry of the system as its corresponding charge operator does not have a well-defined action on the Hilbert space \cite{Cordova:2019jnf}. However, as we shall see below, it is important that $a$ can be thought of as a classical background field for $U(1)\meno_I$ with a well defined field strength.

The second term in \eqref{axion} accounts for the possibility of magnetic monopole configurations in the effective theory. The topological charge of these objects is measured by integrating the time component of $\Tr F \wedge \cD\Phi$. This gives a non-trivial answer since $\Phi$ has a non-trivial winding due to the reduction from five dimensions.

When we reduce the five-dimensional theory on a circle, we get a four-dimensional effective theory which is valid at distances much larger than the radius $\beta$. In particular, the symmetries of the five-dimensional theory descend to four dimensions, and the resulting effective theory is characterized by the following symmetry group
\be\label{effsym}
\mc{G}_{\eff} = U(1)_I\meno \times \bZ\zero_N\times \bZ\one_N\ed
\ee
Here we use $U(1)_I\meno$ simply to denote the transformation property of the axion $a$. We can think about $U(1)\meno_I$ in the effective theory as a special example of non decoupling due to the reduction of the $U(1)_I\zero$ symmetry in five dimensions, that is, the invariance of the partition function under a background $U(1)_I\zero$ gauge transformation in five-dimensions gives rise to a compact scalar field $a$ in the effective theory. In addition to the $\bZ\one_N$ symmetry, the effective theory has a discrete zero-form symmetry $\bZ\zero_N$, which we can think of as coming from the dimensional reduction of the background 2-form gauge field $\cB$ introduced in Section \ref{subsec:CenterSymm}. In thermal physics the order parameter for $\bZ\zero_N$ is a Polyakov loop.

Contrarily to the case of anomalies involving ordinary global symmetries, a mixed anomaly involving a higher-form symmetry persists at finite temperature. By this we mean that anomalies involving higher-form background gauge fields remain non-trivial upon circle reduction. Indeed, consider the dimensional reduction of the anomaly \eqref{eq:6danomaly} 
\be\label{5danomaly}
A_{5}[a, \cB,\cC] = \exp\left( \frac{2\pi \ii}{N}\int_{\cY_5} \frac{\rd a}{2\pi} \,  \cC \, \cB\right) \ec 
\ee
where $\partial \cY_5 = \cM_4$ and $\cC$ is a background gauge field for the $\bZ\zero_N$ symmetry. All the elements of $\cG_\eff$ participate in a mixed 't Hooft anomaly. The above anomaly implies that in the  effective theory all the symmetries \eqref{effsym} \emph{cannot be} simultaneously unbroken in a trivially gapped vacuum. 

We can now see how this anomaly affects the finite temperature behavior of the model. Because of the lack of renormalizability, we shall always assume that the five-dimensional theory is regulated by a UV cutoff scale $\Lambda_5$. In addition to the cutoff scale, we also have a scale $\cO(\beta^{-1})$ associated to the massive Kaluza--Klein modes. It is useful to use $SU(N)$ gauge freedom to diagonalize $\Phi$. For example, when $N=2$, it is common to take $\Phi = \phi_5(x) \sigma_3$ and restrict the discussion to the compact scalar field $\phi_5(x)$.\footnote{When $N > 2$, we could also add a five-dimensional Chern--Simons coupling to the original action, whose dimensional reduction gives 
\be -\frac{\ii k}{24\pi^2}\int_{\cM_4\times S^1_\beta}CS_5(A) = - \frac{ \ii \beta k}{4\pi}\int_{\cM_4} \Tr \Phi \, F_4\wedge F_4 \ed \nonumber\ee
Since $\Phi$ is diagonal, acting with $\bZ\zero_N$ on its compact scalar eigenvalues gives rise to a non-invariant term \beq
- \frac{\ii k}{4\pi N} \int_{\cM_4}\Tr F_4 \wedge F_4 \in - \frac{2\pi \ii k}{N} \Z
\nonumber\ed\eeq This in turn implies that the correct zero-form symmetry in the reduced theory is $\bZ\zero_{\gcd(N,k)}$, confirming the five-dimensional analysis of Section \ref{subsec:YMCS}.}  This field gets a mass contribution from one-loop effects given by $m_{5}^2 \sim g_5^2 \,\beta^{-3}$ (see, for example, \cite{ZinnJustin:2000dr} for a review of thermal field theory).

In the small-$\beta$/high-temperature limit, we can integrate out the adjoint scalar and obtain an effective Lagrangian which is that of pure four-dimensional $SU(N)$ Yang--Mills. At low energies, it is expected that the effective four-dimensional theory confines, so we also expect the one-form $\bZ\one_N$ symmetry to be preserved, whereas we argue that, in the high-temperature limit, $\bZ\zero_N$ should be spontaneously broken. This can be checked by looking at the finite temperature one-loop effective potential for the adjoint scalar which has distinct minima signaling spontaneous breaking. \\
We thus propose that in the high-temperature limit, the vacuum of the four-dimensional effective theory is fully consistent with the anomaly \eqref{5danomaly} when
\begin{equation}
   \begin{matrix} 
  &U(1)_I\meno & \times & \bZ\zero_N & \times& \bZ\one_N  \\
  \textrm{small-}\beta & \checkmark & &\textrm{broken}  &  &\checkmark  \\
  \end{matrix}
\end{equation}

In the large-$\beta$ limit, it is no longer legitimate to integrate out the adjoint scalar. There is a range of values for $\beta$ in which we can still retain a four-dimensional effective description whose Lagrangian is that of pure Yang--Mills with an adjoint scalar. However, as we increase $\beta$ the details of the five-dimensional theory cannot be ignored and we should instead consider an effective five-dimensional theory on a circle. We can again resort to the anomaly \eqref{5danomaly} to propose which symmetries should be preserved by the vacuum at low temperature
\begin{equation}
   \begin{matrix} 
  &U(1)_I\meno & \times & \bZ\zero_N & \times& \bZ\one_N  \\
  \textrm{large-}\beta: & \checkmark & & \checkmark &  &\textrm{broken}  \\
  \end{matrix}
\end{equation}
As we decrease the temperature, the distinct vacua of the effective potential for $\phi_5$ merge into each other, thus we expect $\bZ_N\zero$ symmetry restoration. We do not expect the five-dimensional theory on a circle to have confining behavior, which would be instead related to a fully gapped phase realizing $\bZ\one_N$. One way in which the resulting vacuum is consistent with the anomaly \eqref{5danomaly} is that $\bZ\one_N$ is spontaneously broken. 

In our qualitative analysis of Yang--Mills theory at finite temperature, we assumed that there is no spontaneous breaking of global symmetries at zero temperature. However, there can be different dynamical scenarios all consistent with the mixed anomaly where such possibility cannot be excluded. These will be discussed in Section \ref{scenarios}.

Consider now a different reduction where we \emph{fix} the integral of the background gauge field along the circle to be a constant
\be
\int_{S^1_\beta}\cA = \mu\ec
\ee
that is, we introduce a chemical potential for the instanton symmetry $U(1)\zero_I$. Note that the chemical potential $\mu$ is subject to the identification
\be
\mu \simeq \mu + 2\pi \ed
\ee
In this case, as we consider $\beta \to 0$, the coupling \eqref{eq:coupling} is reduced to a four-dimensional $\theta$ term
\be\label{eq:theta}
\frac{\ii \mu}{8\pi^2} \int_{\Sigma_4} \Tr(F\wedge F)\ec
\ee
with angle $\theta_{4d}=\mu$. As emphasized in \cite{Komargodski:2017dmc}, this confirms that the five-dimensional mixed anomaly between $U(1)\zero_I$ and $\bZ\one_N$  found in \eqref{eq:LargeGaugeBv2} has to exist because it is the uplift of the anomaly discussed in \cite{Gaiotto:2017yup}.

An interesting consequence of the dimensional reduction with a fixed chemical potential for the $U(1)\zero_I$ is that the five-dimensional anomaly is again persistent for any value of the radius circle $\beta$. From this analysis, we immediately see that pure five-dimensional $SU(N)$ Yang--Mills theory at finite temperature is \emph{always} in an ordered phase. Thus, we can analytically confirm and extend certain results regarding the phase diagram of finite temperature Yang--Mills theory obtained in lattice field theory. An early exploration of $SU(2)$ Yang--Mills theory on the lattice was carried out in \cite{Creutz:1979dw}. More recently, the authors of \cite{deForcrand:2010be} have proposed that $SU(2)$ Yang--Mills theory on a circle should be in a certain spontaneously broken phase, which they called ``dimensionally reduced'', for any value of the compactification circle radius. The behavior suggested by lattice analysis matches the predictions made in this section. Moreover, using the anomaly argument, we can also immediately predict that this will also hold true for $N>2$ where lattice results are not yet conclusive.

\subsection{Dynamical Scenarios}
\label{scenarios}

A hallmark of 't Hooft anomalies is that they must match along any RG flow. Often, this gives rise to useful non-perturbative constraints on the dynamics of a given quantum field theory.
In this section we explore what the mixed $U(1)\zero_I \times \bZ\one_N$ anomaly \eqref{eq:6danomaly} implies for pure five-dimensional Yang--Mills theory. Since Yang--Mills theory in $d>4$ is a non renormalizable field theory, we will always assume that a UV cutoff $\Lambda_5$ is present.\footnote{In this section we will restrict our attention to theories without supersymmetry. Certain supersymmetric 5d Yang--Mills theories are expected to be UV completed by a superconformal fixed point. In these examples, the assumption about $\Lambda_5$ can thus be relaxed. We will further discuss supersymmetric theories in Section \ref{sec:SUSYtheories}.} The basic idea is that the mixed 't Hooft anomaly will only give useful constraints about the UV behavior of the theory up to energies $E \lesssim \Lambda_5$. Conversely, if we consider an RG flow {described} by the five-dimensional Yang--Mills Lagrangian, the anomaly must be matched independently of what kind of endpoint such flow might have.

We now discuss a list of candidate phases, all compatible with the anomaly \eqref{eq:6danomaly}, that can be realized dynamically by the UV theory. The mixed anomaly enforces that \emph{none} of these is compatible with a unique trivially gapped vacuum.\footnote{If all the global symmetries that we observe in the IR theory are accidental, anomaly matching will not lead to useful constraints. For the purpose of this section we assume that this is not the case.}

\begin{enumerate}

\item[1.)] \underline{Spontaneous breaking of $\bZ\one_N$}\\
The first option that we consider is that the UV theory saturates the anomaly by spontaneously breaking the one-form symmetry $\bZ\one_N$. The theory is thus both gapless and deconfined. As discussed in the previous section, this scenario is also consistent with anomaly matching at small temperature. We expect that this will be the candidate phase realized by the dynamics. Note that the anomaly could also be consistent with a non-trivial $U(1)\zero_I$-preserving, gapless CFT. It is not at all clear if such conformal field theory exists. So far, the search for interacting conformal fixed points in five spacetime dimensions has not been very successful. As noticed in \cite{Peskin:1980ay}, Yang--Mills theory has a UV fixed point in $d= 4 + \epsilon$ but it is not obvious that such fixed point would survive for $\epsilon = 1$.  

\item[2.)] \underline{Spontaneous breaking of $U(1)\zero_I$}\\
We could also consider the possibility that the one-form symmetry $\bZ\one_N$ is preserved while the global $U(1)\zero_I$ symmetry is spontaneously broken. In this case, there would be a corresponding compact Nambu--Goldstone boson $\chi$ transforming as $\chi \to \chi + \lambda^{(0)}$ under a $U(1)\zero_I$ gauge transformation. However, this is not possible as long as the $5d$ gauge theory description is valid. 
Indeed, in order for $U(1)\zero_I$ to be spontaneously broken the instantonic particles, whose mass is proportional to $g_5^{-2}$, should become massless and this is never possible at weak coupling.

\item[3.)]\underline{Gapless phase preserving both $\bZ\one_N$ and $U(1)\zero_I$}\\
As a third option, we could have a hypothetical gapless CFT which preserves both symmetries $\bZ\one_N$ and $U(1)\zero_I$. For interacting CFTs, the comments in $1.)$ still apply. 

\item[4.)]\underline{Gapped degrees of freedom}\\
The anomaly is of finite order, so the Coleman--Grossman theorem \cite{Coleman:1982yg} does not apply.
Therefore, this is also compatible with a non-trivially gapped topological phase i.e. a topological quantum field theory (TQFT). In the UV, such phase can only be realized together with other local massless degrees of freedom.  These could be thought of as gapped and \emph{deconfined} phases described by a non-trivial five-dimensional $\Z_N$ TQFT.\footnote{Here we are considering a mixed anomaly between a continuous and a finite group. For anomalies involving finite groups, a number of studies exist on the resulting dynamical constraints, including \cite{Wang:2017loc, Tachikawa:2017gyf, Wan:2018djl, Wan:2019oyr, Cordova:2019bsd, Thorngren:2020aph}.}

\end{enumerate}

\section{Comments on Supersymmetric Theories}
\label{sec:SUSYtheories}
\subsection{Pure Super-Yang--Mills Theory}\label{pure}
Pure Yang--Mills theory in five dimensions can be easily embedded into $\mc{N}=1$ (pure) super-Yang--Mills. In addition to Poincaré symmetry, the theory has an $SU(2)_R\zero$ $R$-symmetry acting on the supercharges. Furthermore, since the conservation of the instanton current \eqref{eq:JI} only depends on the Bianchi identity, SYM also enjoys the $U(1)_I\zero$ symmetry. The Lagrangian on $\R^5$ is given by
\beq
\label{eq:FlatSpaceLagrangian}
\SL_{\rm YM} = \frac{1}{g_5^2}\Tr\left( \frac{1}{2}F_{\mu\nu}F^{\mu\nu} + \mc{D}_{\mu}\phi \mc{D}^{\mu}\phi + \ii \overline{\lambda} \gamma^\mu \mc{D}_\mu \lambda - D^i D^i  - \ii {\overline{\lambda}} [\phi ,\lambda ]   \right) \ed
\eeq
The field content above is that of a five-dimensional vector multiplet: the gauge field $A_\mu$, a real scalar $\phi$, two symplectic Majorana spinors  $\lambda^i$ transforming as a doublet of $SU(2)_R\zero$, and three auxiliary real scalar fields $D^i$ in the vector representation of $SU(2)_R\zero$.\footnote{In flat spacetime we can assume that $\overline{\lambda} \equiv \lambda^\dagger$. On a general curved background, the adjoint operation needs to be defined with more care. Note that instanton operators should be properly analyzed by placing the five-dimensional theory on a $S^1\times S^4$ background. A supersymmetric Lagrangian on $S^1\times S^4$ has been proposed in \cite{Nekrasov:1996cz, Kim:2012gu}  (see also \cite{Alday:2015lta, Pini:2015xha}). It is currently not known how to obtain supersymmetric instanton operators on such background except for the point-like (singular) limit.}
All fields transform in the adjoint representation of the gauge group, so we retain the $\bZ\one_N$ symmetry. Therefore, the mixed 't Hooft anomaly $U(1)_I\zero \times \bZ\one_{N}$ derived in \eqref{eq:6danomaly} still holds (and the same is true of the generalization to other gauge groups). Also note that it is possible to add five-dimensional Chern--Simons terms, as in the $SU(N)_k$ case considered in Section \ref{subsec:YMCS}. In that case, we still have a $\bZ\one_{\gcd(N,k)}$ symmetry, as confirmed in \cite{Morrison:2020ool, Albertini:2020mdx, Closset:2020scj, Bhardwaj:2020phs}.  

Matter is organized in hypermultiplets, each containing a pair of dynamical complex scalars and a dynamical symplectic Majorana spinor. A system of $N_f$ hypermultiplets transforms under the symmetry group $Sp(N_f)$, which is also used to impose the reality condition on the component fields, and the coupling to the vector multiplet consists in gauging a subgroup of $Sp(N_f)$. Adding matter in a generic representation may break the one-form center symmetry but it introduces zero-form flavor symmetry.

All five-dimensional $\cN=1$ gauge theories have interesting (real) moduli space of vacua parametrized by the vev $\langle \phi \rangle$ of the vector multiplet scalar. These are called Coulomb branches. In the following, we will only consider the case of gauge group $SU(2)$ and $N_f$ hypermultiplets transforming in the fundamental representation, so the Coulomb branch will be one-dimensional. Letting $\langle \phi \rangle > 0$, the theory is in a Coulomb phase, and the gauge group is broken to $U(1)$. The low-energy effective theory is described by a prepotential, which determines the effective gauge coupling (corresponding to the Coulomb branch metric) as a piecewise linear function of the modulus $\phi$:
\beq
\frac{1}{g^2_{\rm eff}} = \frac{1}{g_5^2} + 8\phi - \frac{1}{2}\sum_{i=1}^{N_f} \left( \left| \phi + m_i \right| + \left| \phi - m_i \right| \right) \, .
\eeq
For the above expression to make sense physically, $g_{\rm eff}^{-2}$ has to be non-negative. Let us now analyze the limit of massless quarks where
\be
 \frac{1}{g^2_{\rm eff}}  =  \frac{1}{g_5^2} + (8-N_f)\phi \ed
\ee 
For $N_f >8$ there is a point in the moduli space where $g_{\rm eff}^{-2}$ would vanish, corresponding to a singularity of the effective theory description. These theories are thus non-renormalizable and require $g^{-2}_5 \neq 0$ as a UV cutoff. The case of $N_f=8$ is rather peculiar since $g_{\rm eff}^{-2} = g^{-2}_5$. As a result, it is impossible to take the strong coupling limit $g_5^{-2} \to 0$ without having the effective coupling degenerate. Finally, for $N_f \leq 7$, $g^{-2}_{\rm eff}$ is positive. Here, the strong coupling limit $g_5^{-2} \to 0$ does not lead to any singular behavior. This strikingly suggests that, at the origin of the Coulomb branch $\phi = 0$, there is a non-trivial five-dimensional scale-invariant field theory \cite{Seiberg:1996bd}. 

Whilst the global symmetry algebra of the low-energy gauge theory is $\mf{u}(1)_I \oplus \mf{so}(2N_f) \oplus \mf{su}(2)_R$, the global symmetry of the UV completion originally found in \cite{Seiberg:1996bd, Douglas:1996xp, Morrison:1996xf} is enhanced to $E_{N_f+1} \oplus \mf{su}(2)_R$.\footnote{Here, $E_1 \equiv \mf{su}(2)$, $E_2 \equiv \mf{su}(2)\oplus \mf{u}(1)$, $E_3 \equiv \mf{su}(2)\oplus \mf{su}(3)$, $E_4 \cong \mf{su}(5)$, $E_5 \cong \mf{so}(10)$, and $E_6, E_7, E_8$ are the exceptional Lie algebras.} Therefore, it is common to use the name $E_{N_f+1}$ to refer to the fixed point from which we flow to the $SU(2)$ gauge theory with $N_f$ fundamental hypermultiplets. This global symmetry enhancement was originally suggested appealing to string-theoretic reasoning, but has since been verified in numerous examples using a variety of methods (see e.g. \cite{Intriligator:1997pq, Aharony:1997bh, Kim:2012gu, Iqbal:2012xm, Bashkirov:2012re, 
Mitev:2014jza, Tachikawa:2015mha, Yonekura:2015ksa, Cremonesi:2015lsa, Gaiotto:2015una, Chang:2017cdx, Chang:2017mxc, Closset:2018bjz, Apruzzi:2019opn}).
It is clear that the way in which five-dimensional gauge theories can be UV completed may not be unique and we make no such claims in this regard.\footnote{Moreover, a criterion to decide whether five-dimensional supersymmetric gauge theories admit a UV completion as a SCFT is not yet available (see e.g. \cite{Jefferson:2017ahm}).}

Arguably the simplest example is that of pure SYM with gauge group $SU(2)\cong Sp(1)$ and vanishing discrete $\theta$ angle (see footnote \ref{footnote:discrete}), for which the UV completion, called $E_1$, has global symmetry algebra $\mf{su}(2)_I\oplus \mf{su}(2)_R$. In particular, the $U(1)_I\zero$ instanton symmetry of the gauge theory is enhanced to (potentially a subgroup of) $SU(2)_I\zero$. In the effective description as $SU(2)$ SYM \eqref{eq:FlatSpaceLagrangian}, there is a mixed anomaly \eqref{eq:6danomaly} between the instanton symmetry $U(1)\zero_I$ and the center symmetry $\Z_2\one$, and its anomaly matching is compatible with the following options.
\begin{enumerate}
\item[1.)] \underline{Spontaneous breaking of $\Z\one_2$}\\
One possibility is that pure $SU(2)$ SYM exhibits spontaneous symmetry breaking of $\bZ\one_2$ and it is described, at high energies, by a gapless deconfined phase. This is perfectly consistent with $E_1$ being the candidate UV fixed point. It is also the natural expectation we could have by looking at the deep IR, where the theory is free and $\Z_2\one$ is spontaneously broken. It is possible to show that this phase saturates the anomaly by shifting the Abelian gauge field in a way analogous to \eqref{eq:Shift}. \item[2.)]\underline{Gapless phase preserving both $\bZ\one_2$ and $U(1)\zero_I$}\\
The mixed 't Hooft anomaly \eqref{eq:6danomaly} is also consistent with a non-trivial SCFT which preserves all the global symmetries. If this is the case, the  UV global symmetry group should at least contain
\be
G_{UV} \supseteq U(1)\zero_I \times \bZ\one_2\ed
\ee
If we assume that the anomaly \eqref{eq:6danomaly} has to be matched by a UV expression, it is interesting to ask whether this might be compatible with global symmetry enhancement. We argue that, in this case, at the UV fixed point $U(1)\zero_I$ enhances to $SO(3)\zero_I$ rather than $SU(2)\zero_I$. The UV anomaly can be written as
\beq
\label{eq:SU2Anomaly}
A_6[\cA,\cB] = \exp\left( \frac{2\pi \ii}{4} \int_{\cY_6}w_2(\cA) \, \cP(\cB) \right)
\eeq
where $w_2(\cA)$ is the second Stiefel--Whitney class of a $SO(3)$ bundle.  Under the symmetry breaking pattern $SO(3)\to SO(2)$ we can write
\be
w_2 = c_1 \mod 2\ec
\ee
and thus match the IR anomaly expression \eqref{eq:6danomaly}. The above matching argument is only compatible with enhancement of the global symmetry group to $SO(3)\zero_I$ rather than $SU(2)\zero_I$. This happens simply because there is no degree-two characteristic class for an $SU(2)$ bundle. Such scenario is compatible with known results. Recall that the five-dimensional superconformal index detects which representations of the global symmetry the states should transform in. Upon direct inspection of the first few orders in the expansion of the index \cite{Nekrasov:1996cz, Kim:2012gu}, we notice that the states transform in representations of $SO(3)$. 
Stronger evidence in support of this claim could be obtained by computing supersymmetric observables that can probe the topology of the global symmetry group as it was done for the $E_{n\geq 2}$ fixed points in \cite{Chang:2016iji}.\footnote{The same discussion holds for $SU(N)_{|N|}$ SYM, where again there is a mixed anomaly between $U(1)_I\zero$ and the center $\Z\one_N$, and the instanton symmetry enhances in the UV to $\mf{su}(2)_I\zero$ \cite{Bergman:2013aca}.} Moreover, starting from the symmetry-preserving fixed point, it is possible to turn on a supersymmetry breaking deformation preserving only a $U(1)_I\zero \times U(1)_R\zero$ subgroup of the enhanced global symmetry. Since the SCFT saturates the anomaly \eqref{eq:SU2Anomaly}, by anomaly matching so should the fixed point of \cite{BenettiGenolini:2019zth}.
\end{enumerate}

Naively, we could expect spontaneous symmetry breaking of the $U(1)\zero_I$ instanton symmetry to occur on the Coulomb branch of the theory, where the gauge group is broken to $U(1)$ and it is not possible to write $J_I$ as in \eqref{eq:JI}. However, spontaneous breaking of $U(1)\zero_I$ does not occur there, even for the theories with fundamental hypermultiplets: we can express the central charge $Z$ of the five-dimensional supersymmetry algebra (in the case with massless hypers) as
\be
\label{eq:CentralCharge}
|Z| = \left| \left( Q_e + 2(8-N_f)Q_I \right) \phi + \frac{Q_I}{g^2_5} \right|
\ee
where $Q_e$ can be thought of as the electric charge in the Cartan of $SU(2)$ and $Q_I$ is (proportional to) the $U(1)\zero_I$ charge. This formula shows that the Coulomb branch description contains BPS instantonic particles charged under $U(1)\zero_I$ with any $Q_I$, which is therefore generically not spontaneously broken there. We will discuss another scenario where spontaneous breaking of $U(1)\zero_I$ takes place at the end of the next section.

We conclude this section with some further comments about spontaneous breaking of the $\bZ\one_2$ symmetry. We consider $Sp(N)$ super Yang--Mills with a hypermultiplet in the rank-2 antisymmetric representation and vanishing $\theta$ angle. The theory has a $Sp(1)\zero\times U(1)_I\zero$ zero-form global symmetry and a $\Z\one_2$ one-form symmetry subject to a mixed 't Hooft anomaly  \eqref{eq:SpNAnomaly}, since adding a hypermultiplet in the rank-2 antisymmetric representation does not break the center symmetry. The conjectured RG fixed point for this model exhibits ordinary global symmetry enhancement to $Sp(1)\zero\times E_1\zero$ \cite{Intriligator:1997pq} and is dual, at large $N$, to the massive IIA AdS$_6$ supergravity background of \cite{Brandhuber:1999np}. The order parameter for $\bZ\one_2$ is the vev of a BPS fundamental Wilson loop, which in the dual picture can be thought of as a fundamental string with endpoints anchored at the boundary of AdS$_6$. Both the field theory and gravity results in \cite{Assel:2012nf} show that BPS Wilson loops obey a perimeter law, indicating that $\bZ\one_2$ is indeed spontaneously broken.

\subsection{Fundamental Matter}
\label{sec:FundamentalMatter}

Adding matter fields in the fundamental representation of the gauge group leads to explicit breaking of the one-form center symmetry. Nevertheless, there might still be a mixed 't Hooft anomaly between the $U(1)_I$ instanton symmetry and the ordinary global symmetry acting on matter fields.\footnote{In this section we avoid writing superscript ${}\zero$ on the groups, as every symmetry is zero-form.} This happens because we can introduce gauge bundles with non-integer instanton number by compensating with the ordinary global symmetry. In four-dimensional QCD, this is studied in the context of color-flavor center symmetry \cite{Cohen:1983sd, Cherman:2017tey, Komargodski:2017smk, Shimizu:2017asf, Gaiotto:2017tne, Tanizaki:2017qhf, Tanizaki:2017mtm, Cordova:2018acb, Cordova:2019uob, Wan:2019oax, Anber:2019nze}. In fact, a similar computation can be applied to five-dimensional Yang--Mills with fundamental quarks, as we show in Appendix \ref{app:5dQCD}. In the following, we focus on the case of $SU(2)$ super-Yang--Mills theory with $N_f$ fundamental hypermultiplets.

The zero-form global symmetry group of the theory is $U(1)_I\times SU(2)_R\times Spin(2N_f)$, corresponding, respectively, to the instanton symmetry, the $R$-symmetry and the flavor symmetry rotating the hypermultiplets. The appearance of $Spin(2N_f)$  rather than $SO(2N_f)$ cannot be seen from the Lagrangian, but can be argued for by studying the fermionic zero-modes in an instanton background \cite{Seiberg:1996bd, Tachikawa:2015mha}.

However, it is important to identify which global symmetry group acts faithfully on the theory. Note that elements in the center of $Spin(2N_f)$ can be identified with elements in the center of the gauge group $SU(2)$, so we focus on
\beq
\label{eq:QuotientNf}
\frac{SU(2)\times Spin(2N_f)}{\Z_2} \ec
\eeq
and, because of the quotient, bundles in $Spin(2N_f)/\Z_2$ may induce $PSU(2)\cong SO(3)$ gauge bundles.

The center of $Spin(2N_f)$ is either $\Z_4$ or $\Z_2\times\Z_2$ depending on whether $N_f$ is, respectively, odd or even, but $Spin(2N_f)/\Z_2\cong SO(2N_f)$ in either cases, by definition. The obstruction to lifting a $SO(r)$ bundle to a $Spin(r)$ bundle is the second Stiefel--Whitney class $w_2\in H^2(BSO(r), \Z_2)$. As discussed in Section \ref{sec:pureYM}, we introduce a background field $\cB\in H^2(\cM_5, \Z_2)$ to keep track of the global symmetry $Spin(2N_f)/\Z_2$, and we set it equal to the lifting obstruction $w_2(\cE_f)$ for the $Spin(2N_f)/\Z_2$ bundle. On the other hand, the quotient \eqref{eq:QuotientNf} relates gauge and global symmetry bundles, so we can tune
\beq
w_2(\cE) = w_2(\cE_f) \, ,
\eeq
and sum over gauge bundles in $SO(3)$ with second Stiefel--Whitney class fixed by the background gauge field for the global symmetry. In turn, as in \eqref{eq:FractionalPartSUN}, the second Stiefel--Whitney class of the gauge bundle uniquely fixes the fractional part of the instanton number. As already seen in Section \ref{sec:pureYM}, the presence of an instanton with fractional $Q_I$ determines a non-trivial transformation of the partition function under a large $U(1)_I$-gauge transformation
\beq
\label{eq:LargeGaugeGlobal}
Z[\cA,\cB] \to Z[\cA,\cB] \exp \left( \frac{2\pi\ii}{4}\ell \int_{\Sigma_4} \cP(\cB) \right) \, .
\eeq
Here, as before, $\ell$ is the winding number of the large gauge transformation around a one-cycle, and $\Sigma_4$ is the base of the fibration. In order to conclude that this transformation signals a mixed anomaly between $U(1)_I$ and the global symmetry group, we should first check that there is no local counterterm that can cancel it. For $N_f\geq 2$, the only relevant local counterterm is
\beq
\label{eq:MatterCounterterm}
\begin{split}
 \exp \left[ 2\pi \ii \, n \int \frac{\cA}{2\pi} \wedge \left( \frac{1}{8\pi^2}\Tr F_{\cE_f}\wedge F_{\cE_f} \right) \right]  \ec
\end{split}
\eeq
normalized so that $n\in \Z$ for a $Spin(2N_f)$ background. However, this can never cancel \eqref{eq:LargeGaugeGlobal}, as the instanton number of a $SO(r)$ bundle with $r\geq 4$ is always integer.\footnote{This follows from Wu's formula on a spin four-manifold \cite[Sec. 6.1]{Aharony:2013hda}.} Therefore, we conclude that there is a mixed 't Hooft anomaly between $U(1)_I$ and $Spin(2N_f)$ for $N_f\geq 2$, with anomaly theory
\be\label{eq:6danomalyFlavor}
A_6[\cA, \cB] = \exp\left( \frac{2\pi \ii}{4}\int_{\cY_6} \frac{\rd \cA}{2\pi} \, \cP(\cB) \right) \ed 
\ee
The same reasoning goes through for $N_f=1$, except for the form of the counterterm \eqref{eq:MatterCounterterm}, which in that case would read $c_1(\cE_f)^2$, and would still not cancel the anomalous variation.

We will now briefly comment on the mixed 't Hooft anomaly \eqref{eq:6danomalyFlavor}. As already mentioned, $SU(2)$ SYM with $N_f\leq 7$ hypermultiplets in the fundamental representation has a UV completion to the fixed point $E_{N_f+1}$ with global symmetry group given by the product of the $R$-symmetry $SU(2)_R$ and the simply connected group with algebra $E_{N_f+1}$, representing the enhancement of the IR symmetry  $U(1)\times Spin(2N_f)$.\footnote{The topology of the global symmetry group of the $E_{N_f+1}$ fixed point has been investigated in \cite{Chang:2016iji} using the superconformal index decorated by ``ray operators''.} The presence of an interacting SCFT preserving all symmetries is fully consistent with the anomaly matching. However, it is also possible that some (or all) of the global symmetries are spontaneously broken.

This does not happen on the Coulomb branch of the $E_{N_f+1}$ theory: from the expression for the central charge \eqref{eq:CentralCharge} we see that the instanton symmetry is not spontaneously broken on the Coulomb branch. Nonetheless, $Spin(2N_f)$ can be explicitly broken to $U(1)^{N_f}$ by turning on quarks masses. 

On the other hand, the presence of matter allows us to consider additional moduli space of vacua parametrized by the vevs of the scalar fields in the hypermultiplets, the so-called Higgs branches. These are complex manifolds, more precisely, hyperK\"ahler cones. Whilst the Coulomb branch of the $E_{N_f+1}$ fixed point is the same as that of the $SU(2)$ SYM with $N_f$ hypermultiplets, namely $\R_+$, the Higgs branch can instead vary dynamically. This is due to instantons becoming massless (their mass is proportional to $g_5^{-2}$) \cite{Seiberg:1996bd, Douglas:1996xp, Morrison:1996xf}. Classically, the Higgs branch is given by the reduced moduli space of one $SO(2N_f)$ instanton. At strong coupling, a string theory argument suggests that the Higgs branch is given instead by the reduced moduli space of one $E_{N_f+1}$ instanton on $\C^2$. It was argued in \cite{Cremonesi:2015lsa} that on the Higgs branch of the SCFT, the global symmetry $E_{N_f+1}$ (and its $U(1)_I$ subgroup) is spontaneously broken due to supersymmetric states which are explicitly charged under the instanton symmetry. 

Finally, we would like to comment on an alternative method to saturate the anomaly \eqref{eq:6danomaly} in \emph{pure} SYM theory. Even though there is no matter, in the SCFT there is a protected subsector of $1/2$-BPS operators that generates the coordinate ring of the Higgs branch hyperK\"ahler cone.\footnote{In analogy with four-dimensional $\cN=2$ SCFTs, these can be thought of as Higgs branch operators since $1/2$-BPS operators of 5$d$ SCFTs always have a $SU(2)_R$ quantum number.} Thus, we say that the Higgs branch geometry for the $E_1$ SCFT is a cone $\C^2/\Z_2$. From a gauge theory point of view, this is due to massless instantons in the spectrum opening up two flat directions \cite{Cremonesi:2015lsa}. On this moduli space, the full global symmetry group $E_1$ (and its subgroup $U(1)_I$) is spontaneously broken. Such symmetry breaking is clearly consistent with the IR anomaly discussed in Section \ref{pure}.

\bigskip
\begin{center}
\textbf{Acknowledgments}
\end{center}
\noindent We thank Zohar Komargodski and David Tong for helpful discussions and comments on a draft. We have also benefited from discussions with Fabio Apruzzi, Lakshya Bhardwaj, Stefano Cremonesi, Chris Kusch, Jihwan Oh, Lorenzo Ruggeri, Sakura Sch\"afer-Nameki, Adar Sharon, Carl Turner and Yifan Wang. The work of PBG has been supported by the STFC consolidated grant ST/P000681/1. L.T. is supported in part by the Simons
Foundation grant 488657 (Simons Collaboration on the Non-Perturbative Bootstrap) and
the BSF grant no. 2018204. Any opinions, findings, and conclusions or recommendations expressed in this material are
those of the authors and do not necessarily reflect the views of the funding agencies.

\bigskip

\appendix

\section{Five Dimensional Yang--Mills Theories with Matter}
\label{app:5dQCD}

In this appendix, we add fermions to the Lagrangian \eqref{eq:YM} by introducing $N_f$ massive Dirac spinors transforming in the fundamental of the $SU(N)$ gauge group, thus explicitly breaking the center symmetry $\Z_N\one$.\footnote{As in Section \ref{sec:FundamentalMatter}, from now on we avoid writing the superscript $\zero$, as all symmetries are zero-form.} Nevertheless, as we saw in Section \ref{sec:SUSYtheories}, there still is a mixed 't Hooft anomaly between the instanton symmetry and the global symmetry. The computation mirrors closely the analogous center-flavor symmetry anomaly in four-dimensional QCD \cite{Benini:2017dus, Gaiotto:2017tne, Cordova:2019uob, Anber:2019nze}, and we include it for completeness.

\medskip

The spinor part of the Lagrangian reads
\beq
\SL_{\rm Dirac} = \ii \psi_j^\dagger \slashed{D}\psi^j + m \psi_j^\dagger \psi^j \ec
\eeq
where $\psi^j\in \bC^4$ is a Dirac spinor in five dimension and $j=1, \dots, N_f$. The group acting on the fermions is $SU(N)\times U(N_f)$. However, within the $U(1)$ center of $U(N_f)$ there is a $\bZ_N$ subgroup that can be used to undo a $\bZ_N$ transformation in the center of the gauge group. Therefore, the group that acts faithfully is
\beq
\label{eq:GlobalSymmetryQuotientQCD}
\frac{SU(N)\times U(N_f)}{\bZ_N} \, ,
\eeq
and, because of the quotient, a $PU(N_f)$ bundle for the global symmetry may induce a $PSU(N)$ gauge bundle.

\medskip

First, we notice that in a $U(N_f)$ bundle with connection $F_f$, the $\mf{u}(1)$ part of the connection $\rd \cC= \frac{1}{N_f}\Tr F_f$ is not required to have quantized fluxes: the fractional flux can be compensated by the lifting obstruction of a $PSU(N_f)$ bundle $\cE_f$, the Brauer class $w\in H^2(BPSU(N),\Z_{N_f})$, so that\footnote{This may be more familiar to the reader in the context of $SO$ bundles lifted to $Spin^c$ bundles using an additional ``$U(1)$ bundle'' with fractional fluxes. Indeed, the two coincide for $N=2$, since $PSU(2)\cong SO(3) \cong PU(2)$ and $U(2)\cong Spin^c(3)$, and the Brauer class above reduces to the Stiefel--Whitney class.}
\beq
\label{eq:FractionalInsideUN}
\int \left(\frac{\rd \cC}{2\pi} - \frac{1}{N_f}w ( \cE_f ) \right) \ \in \ \Z \, .
\eeq
However, the quotient \eqref{eq:GlobalSymmetryQuotientQCD} means that the global symmetry of the theory acting faithfully is $U(N_f)/\Z_N$, which we represent as the quotient of $SU(N_f)\times U(1)$ by the equivalence relation
\beq
SU(N_f)\times U(1) \ni (g,\lambda) \sim (g, \e^{-2\pi \ii/N}\lambda) \sim (\e^{2\pi \ii/N_f}g, \e^{-2\pi \ii /N_f}\lambda) \ec
\eeq
modifying \eqref{eq:FractionalInsideUN} into
\beq
\label{eq:ConstraintQuotient}
\int\left( N \frac{\rd\cC}{2\pi} - \frac{N}{N_f}w (\cE_f) \right) \ \in \ \Z \ed
\eeq
From this, we may define the obstruction to lifting a $U(N_f)/\Z_N$ bundle to a $U(N_f)$ bundle as
\beq
\label{eq:ObstructionUN}
\int w^{(N)} \equiv \int\left( N \frac{\rd\cC}{2\pi} - \frac{N}{N_f}w (\cE_f) \right) \ \mod N \, .
\eeq
We now keep track of the global symmetry $U(N_f)/\bZ_N$ using a $\Z_N$ background field $\cB\in H^2(\cM_5, \Z_N)$ that we tune to be equal to the obstruction $w^{(N)}$.
Since the quotient \eqref{eq:GlobalSymmetryQuotientQCD} relates gauge and global symmetry, we may construct $PSU(N)$ gauge bundles $\cE$ cancelling the cocycle obstruction using $w^{(N)}$. That is, we can turn on gauge bundles with fractional instanton number provided the two obstructions are related by
\beq
\label{eq:ObstructiongaugeQCD}
\int w(\cE) = \int w^{(N)} \ \mod N \, .
\eeq
This seems to imply again the existence of an anomaly between $U(1)\zero_I$ and the global symmetry. The form is again \eqref{eq:LargeGaugeBv2}
\beq
\label{eq:LargeGaugeQCD}
Z[\cA,\cB] \to Z[\cA,\cB]\exp \left( -\frac{2\pi\ii}{2N} \ell \int_{\Sigma_4}\cP(\cB) \right) \, ,
\eeq
but now $\cB$ is not anymore a background gauge field for a one-form symmetry. This looks like a finite-order mixed 't Hooft anomaly between two global zero-form symmetries, but a putative anomaly is such only if there is no local counterterm that can cancel it.

In order to make contact with the previous literature, let
\beq
K \equiv \lcm (N,N_f) \, , \qquad L \equiv \gcd(N,N_f) = \frac{NN_f}{K} \, ,
\eeq
and introduce the $U(1)$ gauge field $\tilde{\cC}\equiv K \cC$, with fluxes quantized in integer multiples of $2\pi$. Then we may write \eqref{eq:ObstructiongaugeQCD} in the equivalent way
\beq
\label{eq:ObstructionU1}
\int \frac{\rd\tilde{\cC}}{2\pi} = \int \left( \frac{N_f}{L}w(\cE) + \frac{N}{L}w(\cE_f) \right) \mod K \, .
\eeq
The relevant local counterterms modify the partition function into
\beq
\label{eq:CountertermsQCD}
\begin{split}
Z'[\cA, \cB] 
&= Z[\cA, \cB]  \exp \left[ 2\pi \ii \int_{\Sigma_4} \left( - \frac{s}{2N_f}\mc{P}(w(\cE_f)) + \frac{t}{2}  \frac{\rd\cC}{2\pi}\wedge \frac{\rd\cC}{2\pi} \right)  \right] \ec
\end{split}
\eeq
with terms normalized so that $s,t$ are integers in the case of true $SU(N_f)$ and $U(1)$ bundles.

These counterterms can cancel the anomalous variation of the partition function if
\beq
\exp \bigg[ 2\pi \ii \int_{\Sigma_4} \bigg( \frac{\ell N-t}{2N^2}\mc{P}(w^{(N)}) + \frac{sN_f - t}{2N_f^2}\mc{P}(w(A)) - \frac{t}{NN_f} w(^{(N)})\cup w(A)  \bigg) \bigg] = 1 \, ,
\eeq
which holds iff we can find $\ell ,s,t$ such that
\beq
\begin{split}
\ell N-t \in N^2\Z \, , \qquad sN_f - t \in N_f^2\Z \, , \qquad t\in NN_f\Z \, .
\end{split}
\eeq
These conditions can only be solved if $\ell = 0 \mod L$.
Therefore, they can always be solved iff $L=\gcd(N,N_f)=1$. Otherwise, there will be large gauge transformations for which these cannot be solved, thus potentially leading to an anomaly between the $U(1)_I$ and $\Z_N$. As a matter of fact, the counterterms can be used to show that the anomaly is only with $\Z_L$, since we can use \eqref{eq:ObstructionUN} to write the anomaly in \eqref{eq:LargeGaugeQCD} as
\be
\begin{split}
& \exp\left[ \frac{2\pi \ii}{L}\ell \int_{\Sigma_4}\left( \frac{r}{2} - \frac{L+Nr}{2N} \right) \cP( \cB ) \right] 
= \exp\left[ \frac{2\pi \ii}{L}\ell \int_{\Sigma_4}\left( \frac{r}{2}  \cP(\cB) + j \, w(\cE_f)\cup \cB \right)\right]\\& \times\exp \left[ 2\pi \ii \ell \int_{\Sigma_4} \left( \frac{j N}{2N_fL}\mc{P}(w(\cE_f)) - \frac{jK}{2} \frac{\rd\cC}{2\pi}\wedge \frac{\rd\cC}{2\pi} \right) \right]\ec
\end{split}
\ee
where $j,r\in\Z$ such that $L+Nr=jK$. Thus we can choose $s=j \frac{N}{L}$ and $t=jK$ (both integers) in \eqref{eq:CountertermsQCD} and reduce the $\Z_N$ anomaly to a $\Z_L$ anomaly, with anomaly theory
\beq
A_6[\cA,\cB] = \exp \left( \frac{2\pi\ii}{L}\int_{\cY_6} \frac{\rd\cA}{2\pi} \cup \left( \frac{r}{2}  \cP(\cB) + j \, w(\cE_f)\, \cB \right)\right) \, ,
\eeq
which is independent of the choice of $j, r$ satisfying the constraint $L+Nr=jK$.

\newpage

\renewcommand\refname{\bfseries\large\centering References\\ \vspace{-0.4cm}
\addcontentsline{toc}{section}{References}}

\bibliographystyle{utphys.bst}
{\small
\bibliography{Anomaly5D.bib}%
}
	
\end{document}